\documentclass{article}
\usepackage{arxiv}

\usepackage{lipsum}
\usepackage{amsfonts}
\usepackage{graphicx}
\usepackage{epstopdf}
\usepackage{algorithmic}
\usepackage{amsmath}
\usepackage{xcolor}                  
\usepackage{algorithm}               
\usepackage{subcaption}
\usepackage{mathtools}
\usepackage{bm}
\usepackage{appendix}
\usepackage{multirow}
\usepackage{hyperref}
\usepackage{cleveref}
\usepackage{footnote}
\usepackage{float}

\graphicspath{ {./plots/} }
\numberwithin{equation}{section} 

\definecolor{salmon}{RGB}{255,145,164}

\newcommand{\bx}{\boldsymbol{x}}

\newcommand{\by}{\boldsymbol{y}}
\newcommand{\bq}{\boldsymbol{q}}
\newcommand{\bQ}{\boldsymbol{Q}}

\definecolor{red}{HTML}{D83E29}
\definecolor{green}{HTML}{438F8B}
\definecolor{blue}{HTML}{1D567A}
\definecolor{grey}{HTML}{C8C2B9}
\let\figref=\ref
\renewcommand{\figref}[1]{Figure \ref{#1}}
\let\eqref=\ref
\renewcommand{\eqref}[1]{Equation \ref{#1}}

\newcommand{\Secref}[1]{Section \ref{#1}} 

\newcommand{\E}{\mathrm{E}}
\newcommand{\Var}{\mathrm{Var}}

\ifpdf
  \DeclareGraphicsExtensions{.eps,.pdf,.png,.jpg}
\else
  \DeclareGraphicsExtensions{.eps}
\fi

\usepackage{enumitem}
\setlist[enumerate]{leftmargin=.5in}
\setlist[itemize]{leftmargin=.5in}

\usepackage{amsopn}



\graphicspath{{plots/}}

\newcommand*{\email}[1]{\href{mailto:#1}{\nolinkurl{#1}} } 

\title{Multi-objective optimisation using expected quantile improvement for decision making in disease outbreaks}

\begin{document}
\author{Daria Semochkina\thanks{University of Southampton, UK 
  (\email{d.semochkina@soton.ac.uk}, \email{d.woods@soton.ac.uk})}
\and Alexander I. J. Forrester\thanks{Nottingham Trent University, UK (\email{alexander.forrester@ntu.ac.uk})}
\and David C Woods\footnotemark[2]
}
\maketitle

\begin{abstract}

Optimization under uncertainty is important in many applications, particularly to inform policy and decision making in areas such as public health. A key source of uncertainty arises from the incorporation of environmental variables as inputs into computational models or simulators. Such variables represent uncontrollable features of the optimization problem and reliable decision making must account for the uncertainty they propagate to the simulator outputs. Often, multiple, competing objectives are defined from these outputs such that the final optimal decision is a compromise between different goals. 

Here, we present emulation-based optimization methodology for such problems that extends expected quantile improvement (EQI) to address multi-objective optimization. Focusing on the practically important case of two objectives, we use a sequential design strategy to identify the Pareto front of optimal solutions. Uncertainty from the environmental variables is integrated out using Monte Carlo samples from the simulator. Interrogation of the expected output from the simulator is facilitated by use of (Gaussian process) emulators. The methodology is demonstrated on an optimization problem from public health involving the dispersion of anthrax spores across a spatial terrain. Environmental variables include meteorological features that impact the dispersion, and the methodology identifies the Pareto front even when there is considerable input uncertainty. 

\end{abstract}


\section{Introduction}
\label{subsec:Optimisation:Introduction}

Many applications of product, system or process design, including for public health interventions, require solutions to optimization problems that have multiple, usually competing, goals. Assuming the goal of simultaneous minimization, such a problem may be described as:
\begin{equation}\label{opt_prob}
\begin{array}{rll}
\min & f_i(\bx_c) & \mbox{for } i=1,\ldots, n\\
\text{subject to: }
& \mbox{lb}_k\leq x_{c,k} \leq \mbox{ub}_k & \mbox{for } k=1,\ldots,v,
\end{array}
\end{equation}
where $f_i$ are $n$ black-box objective functions, $\bx_c = (x_{c,1}, \ldots, x_{c,v})^{\sf T}$ represent the values of $v$ controllable variables, 
and $\mbox{lb}_k$ and $\mbox{ub}_k$ are bounds on the input $x_{c,k}$. In addition, various constraints might be present, restricting the solution space.

We focus on multi-objective optimization on a Pareto front \cite{fonseca1995aoo}, a collection of optimal solutions for each $f_i$ such that no single objective can be improved without making at least one of the remaining objectives worse \cite{giagkiozis2014pareto}. We assume the computational expense of the functions $f_i$ is such that only a limited number of evaluations are possible (without corresponding evaluation of derivatives). Hence, we adopt Bayesian optimization methodology \cite{frazier2018bayesian} that assumes no particular knowledge about the functions $f_i$ (e.g.,~convexity or linearity \cite{boyd2004convex,bertsimas1997introduction}) except continuity and (some degree of) smoothness with respect to $\bx_c$.

In disease-outbreak modeling and decision making \cite{walters2018modelling,fischhoff2015realities}, there will inevitably be one dominant objective: that of minimizing the number of fatalities. However, accepting that achieving zero fatalities may be impossible in practice, other factors will come into consideration when assessing decision strategies. In particular, resources, such as a limited supply of vaccines, should be utilized to their maximum efficiency. Formally, these quantities could be treated simply as constraints (e.g.,~minimize fatalities, subject to fewer than 1\% of the population being vaccinated per day) or as additional objectives (e.g.,~simultaneously minimize both fatalities and the rate of vaccination). The latter approach results in a more difficult optimization problem, which becomes less tractable with each additional objective that is included due to the need to measure potential `improvements' in a more complex and higher dimensional space. However, the latter approach is also potentially more informative; in particular, the resulting set of solutions can be examined by decision-makers to better understand the trade-off between objectives \cite{lotov2004interactive,grierson2008pareto}. 

In most disease scenarios, it is also impossible to predict precisely how the outbreak will develop due to unknown or uncertain parameters and/or the stochastic nature of the modeling. In particular, outbreak progression may be uncertain because of dependence on some environmental variables whose values will be unknown. Following advice from subject matter experts, we assume interest is in minimization of expected values $f_i(\bx_c) = \E_{X_e}\left[h_i(\bx_c,\bx_e)\right]$ where $h_i(\cdot, \cdot)$ is a function quantifying an aspect of system performance for given values of the controllable $\bx_c$ and environmental $\bx_e$ variables. The environmental variables can be controlled when evaluating each $h_i(\cdot,\cdot)$ but will be uncontrollable in the physical process that is being modeled. Hence the focus is on optimization with respect to $\bx_c$. The environmental variables are assumed to follow a known distribution that would typically be either elicited from subject matter experts or estimated from available data. In an outbreak model the environmental variables represent features such as outbreak source or relevant meteorological conditions. Optimization of expected values has been subject to recent development \cite{jl2013,tf2022} and also found wide application in a variety of areas \cite{sm2010, lkm2015}.

We develop a multi-objective \cite{forrester2008edv} Bayesian optimization approach to the problem described above, extending the methodology of expected quantile improvement \cite{picheny2013qbo} (\Secref{sec:Optimisation:EQI}). An illustrative example is used to explore the performance of the approach (\Secref{sec:Optimsation:illustrative_example}) and demonstrate its application to the problem of optimizing the response to an anthrax outbreak (\Secref{sec:Optimisation:anthrax}) where we also incorporate constraint functions that reduce the solution space. A discussion of practical policy implications and future research directions are also presented (\Secref{sec:Optimisation:Summary}). Code is publicly available through GitHub \cite{R:MOEEQI}.

\section{Multi-objective emulation-based optimization for noisy functions}
\label{sec:Optimisation:EQI}

We assume Monte Carlo estimates,
\begin{equation}\label{eq:mc_avg}
y_i(\bx_c) = \frac{1}{N} \sum_{r = 1}^Nh_{i}(\bx_c,\bx_e^r)\,, \quad \bx_e^1, \ldots, \bx_e^N\sim X_e\,,
\end{equation}
are available as approximations to the expected values $f_i(\bx_c) = \E_{X_e}\left[h_i(\bx_c,\bx_e)\right]$ ($i =1, \ldots, n$). For large $N$ \cite{cochran1977sampling}, the central limit theorem implies $y_i(\bx_c) \sim \mathcal{N}\left(f_i(\bx_c), \sigma_i^2(\bx_c) / N\right)$ or, equivalently,
\begin{equation}\label{eq:statmodel}
y_i(\bx_c) = f_i(\bx_c) + \varepsilon_i(\bx_c, N)\,,\quad \varepsilon_i(\bx_c, N) \sim \mathcal{N}\left(0, \sigma_i^2(\bx_c) / N\right)\,,
\end{equation}
with $\sigma_i^2(\bx_c) = \Var_{X_e}\left[h_i(\bx_c, \bx_e)\right]$. Hence, the presence of environmental variables results in the availability of only noisy evaluations, $y_i(\bx_c)$, of  the objective function $f_i(\bx_c)$. 

Broadly speaking, Bayesian optimization methods for a single objective function proceed by (i) assuming a prior for the unknown function $f_i(\cdot)$; (ii) selecting new points $\bx_c$ at which to evaluate $f_i(\cdot)$ according to some infill criterion that maximizes an acquisition function; and (iii) updating an estimate of the function minimum, and its location, using the updated posterior for $f_i(\cdot)$. The most common prior for $f_i(\cdot)$ is a Gaussian process (GP) \cite{williams2006gaussian}
$$
f_i(\bx_c) \sim \mbox{\textit{GP}}\left\{\mu_i(\bx_c), \kappa_i(\bx_c, \bx_c^\prime)\right\}\,,
$$
a stochastic process defined via a mean function $\mu_i(\cdot)$ and covariance function $\kappa_i(\bx_c, \bx_c^\prime)$. For any finite set of input vectors $\bx_c^1, \ldots, \bx_c^S$, arranged in a design matrix $X_S = \left(\bx_c^1, \ldots, \bx_c^S\right)^{\mathrm{T}}$ the resulting function outputs follow a multivariate normal distribution
$$
\begin{pmatrix}
f_i(\bx_c^1) \\
\vdots \\
f_i(\bx_c^S)
\end{pmatrix}
\sim \mathcal{N}\left(\boldsymbol{\mu}_i(X_S), K_i(X_S)\right)\,,
$$
with mean vector $\boldsymbol{\mu}_i(X_S)$ having $j$th entry $\mu_i(\bx_c^j)$ and covariance matrix $K_i(X_S)$ having $jk$th entry $\kappa_i(\bx_c^j, \bx_c^k)$ ($j, k = 1, \ldots, S$). 

It is common to assume a linear form for the mean function, with $\mu_i(\bx_c) = g_i(\bx_c)^{\sf T}\boldsymbol{\beta}_i$. In this paper, we assume a constant mean function with $g_i(\bx_c)^{\sf T} = 1$ and $\boldsymbol{\beta}_i = \beta_{i, 0}$. Throughout, we assume a squared exponential covariance function \cite{swn2019}, suitable for smooth response functions, although there is nothing in the methodology specific to this choice. Popular alternatives include the Mat{\`e}rn function, suitable for functions displaying less smoothness.

Assuming known variances $\sigma_i^2(\bx_c^j)$, the GP prior is conjugate to the normal likelihood conditional on any hyperparameters involved in the covariance function. Further, assuming $\beta_{i, 0}\sim \mathcal{N}(0, \tau^2)$ allows analytical marginalization with respect to the constant GP mean; we assume a non-informative prior by allowing $\tau^{-2}\rightarrow 0$. Hence the posterior for $f_i(\bx_c)$, conditional on noisy data $\by_i^S = \left[y_i(\bx_c^1), \ldots, y_i(\bx_c^S)\right]^{\sf T}$ from model~\ref{eq:statmodel}, is also a GP:
\begin{equation}\label{eq:gp-post}
f_i(\bx_c) \,|\, \by_i^S \sim \mbox{\textit{GP}}\left\{m_i(\bx_c), s_i(\bx_c, \bx_c^\prime)\right\}\,,
\end{equation}
with updated mean and covariance functions:
$$
m_i(\bx_c) = \hat{\beta}_{i, 0} + k_i(\bx_c, X_S)\left(K_i(X_S) + \Delta_i\right)^{-1}\left(\by_i^S - \hat{\beta}_{i, 0}\boldsymbol{1}_S\right)\,,
$$
\begin{multline*}
s_i(\bx_c, \bx_c^\prime) = \kappa_i(\bx_c, \bx_c^\prime) - k_i(\bx_c, X_S)^{\sf T}\left(K_i(X_S) + \Delta_i\right)^{-1}k_i(\bx_c^\prime, X_S)\\ 
+ \frac{\left[1 - \boldsymbol{1}_S^{\sf T}\left(K_i(X_S) + \Delta_i\right)^{-1}k_i(\bx_c^\prime, X_S)\right]^2}{\boldsymbol{1}_S^{\sf T}\left(K_i(X_S) + \Delta_i\right)^{-1}\boldsymbol{1}_S}\,,
\end{multline*}
where 
$$
\hat{\beta}_{i, 0} = \frac{\boldsymbol{1}_S^{\sf T}\left(K_i(X_S) + \Delta_i\right)^{-1}\by_i^S}{\boldsymbol{1}_S^{\sf T}\left(K_i(X_S) + \Delta_i\right)^{-1}\boldsymbol{1}_S}\,,
$$ 
$k_i(\bx_c, X_S) = \left[\kappa_i(\bx_c, \bx_c^1), \ldots, \kappa_i(\bx_c, \bx_c^S)\right]^{\sf T}$ and $\Delta_i = \mbox{diag}\left\{\sigma_i^2(\bx_c^1), \ldots, \sigma_i^2(\bx_c^S)\right\}$. These equations are often referred to as the stochastic kriging equations \cite{ans2010}, with $m_i(\bx_c)$ being the best (with minimum mean squared error) linear unbiased predictor for $f_i(\bx_c)$.

Any hyperparameters in $\kappa_i(\cdot, \cdot)$ can be estimated using, e.g., maximum likelihood or cross-validation. If the variances are unknown, they can be estimated from the Monte Carlo sample, 
\begin{equation}\label{eq:mcerror}
\hat{\sigma}^2_{i, N}(\bx_c) = \frac{1}{N-1}\sum_{r = 1}^N\left[h_i(\bx_c, \bx_e^r) - y_i(\bx_c)\right]^2\,,
\end{equation}
and plugged into $\Delta_i$. More sophisticated approaches could also be adopted including joint modeling of the mean and variance of $y_i(\bx_c)$ \cite{bgl2018}. Later in this section, we derive closed-form expectations of improvement functions and areas under the Pareto front that require the model from~\eqref{eq:gp-post} to follow a Gaussian process. Hence, plug-in estimates of $\kappa_i(\cdot,\cdot)$ and $\sigma_i^2$ are required.

\subsection{Gaussian process-based optimization with noisy observations}\label{sec:BO}
\begin{figure}
\begin{subfigure}{.49\textwidth}
\centering
\scalebox{1.7}{\includegraphics[width=.414\textwidth
]{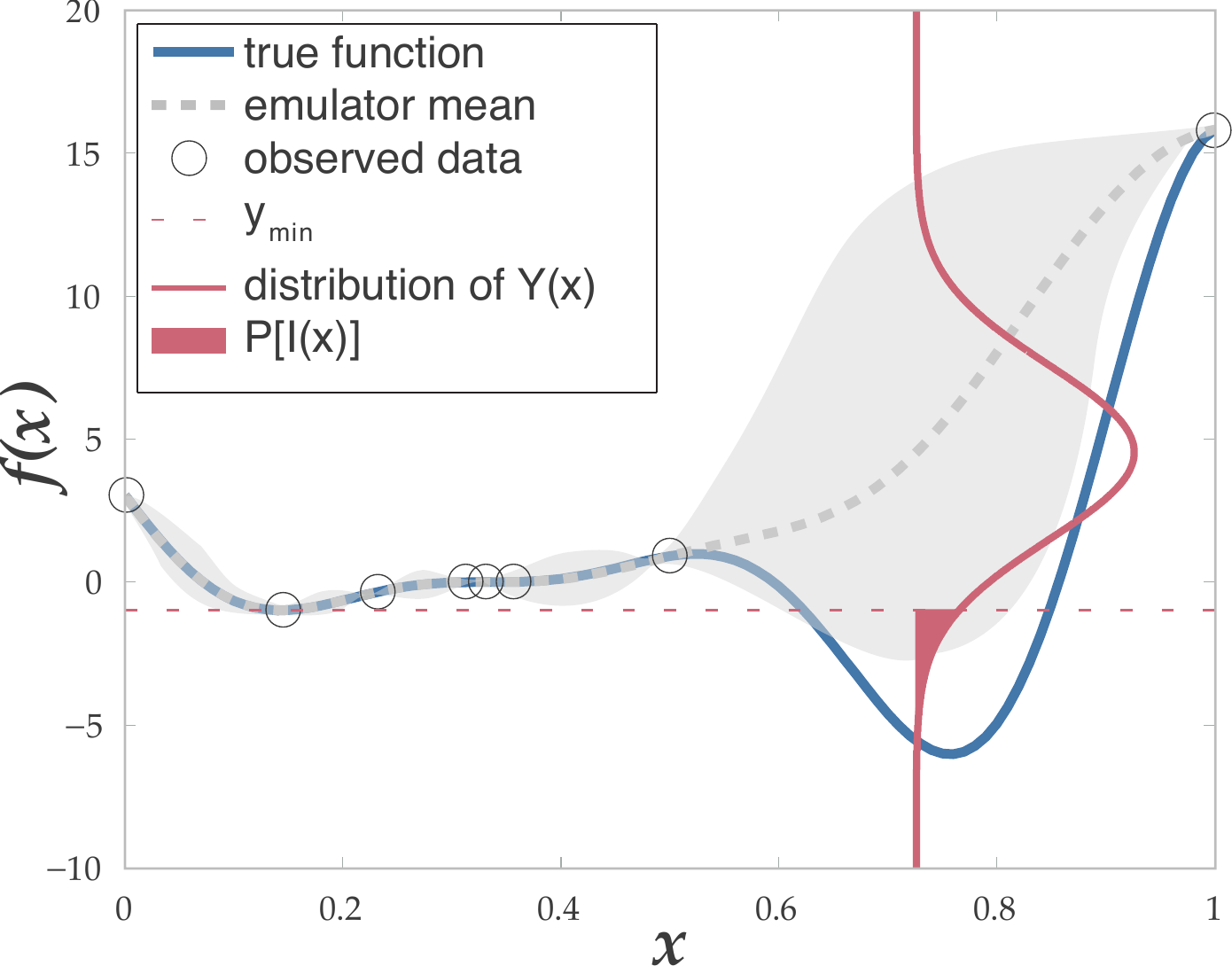}}
\caption{}
\label{fig:EI}
\end{subfigure}
\begin{subfigure}{.49\textwidth}
\centering
\scalebox{1.7}{\includegraphics[width=.586\textwidth
]{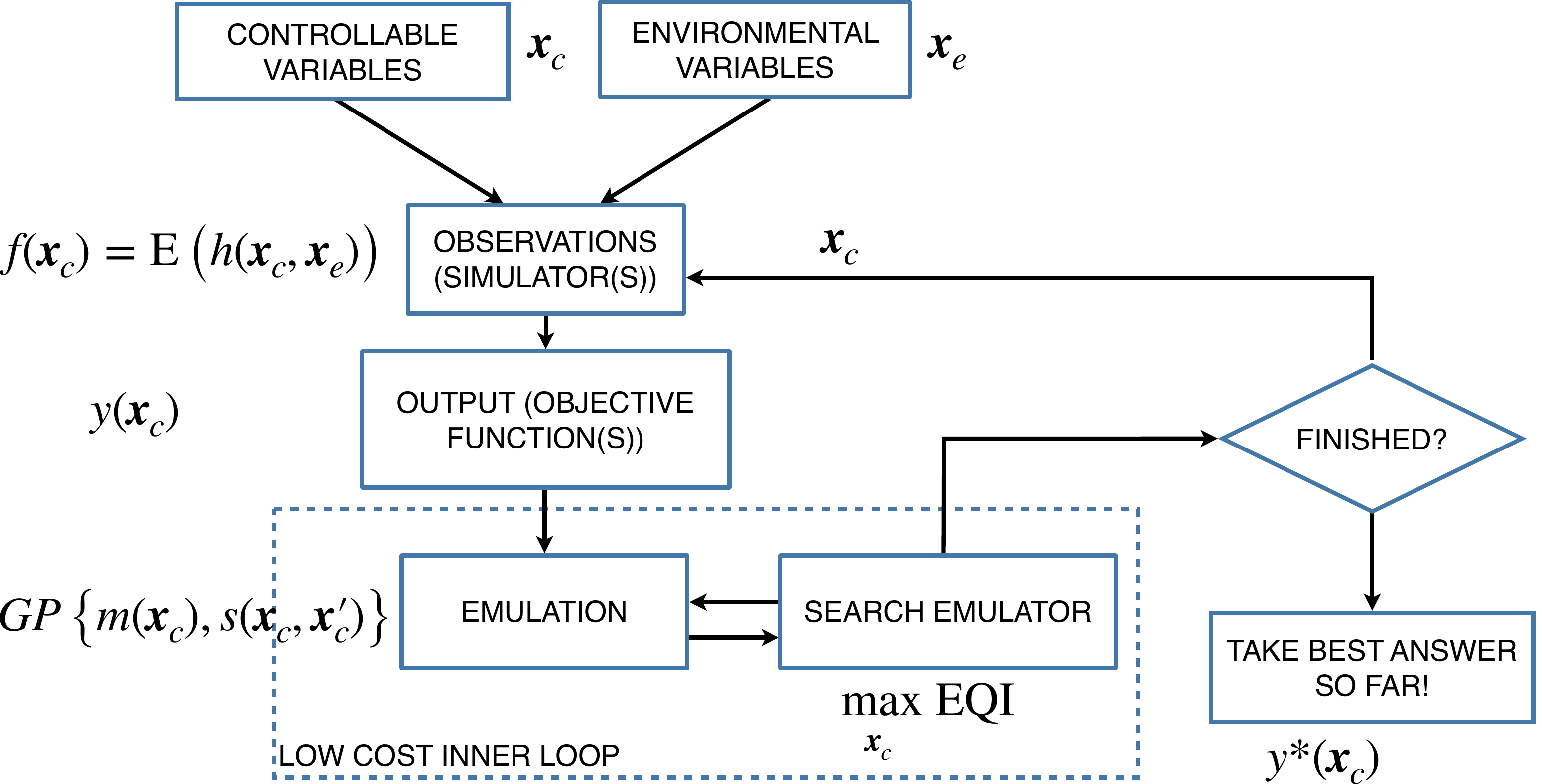}}
\caption{}
\label{opt:envir}
\end{subfigure}%

\caption{(a) Schematic of emulator-based optimization. Image adapted from Forrester et al. \cite{forrester2008edv} (b) Flow-chart of emulation-based optimization with environmental variables.}
\end{figure}

Selection of the next point $\bx_c^{S+1}$ in an optimization scheme for a single function\footnote{For simplicity, we suppress the indexing across the multiple objective functions in this section.} is typically the result of a trade-off between exploiting current knowledge about the minimum of $f(\cdot)$ via the kriging mean $m(\bx)$ and exploring regions where prediction uncertainty is high, as measured by the kriging variance $s^2(\bx_c) = s(\bx_c, \bx_c)$; see Figure~\ref{fig:EI}. For black-box optimization of a deterministic function, the most common criterion for sequential selection of design points is expected improvement (EI) \cite{mtz1978}, the main building block of the popular efficient global optimization algorithm \cite{jones1998ego}. As illustrated in Figure~\ref{fig:EI}, EI depends on the probability of making an improvement over the current minimum and the conditional expectation of that improvement.

However, standard applications of EI assume the function $f(\cdot)$ is directly observable ($\sigma^2(\bx_c) = 0$), using the minimum of $\by^S$ as the current estimate of the minimum possible response and calculating the improvement between this quantity and a deterministic future response. With noisy data these two properties are undesirable however the noise arises, and may lead to an inefficient selection of points and a lack of convergence to the true minimum. In particular, replication of points will offer no expected improvement in this case, whereas it may be highly beneficial for noisy responses.

Probably the most principled extension to address noisy data is the expected quantile improvement (EQI) \cite{picheny2013qbo}, defined as
$$
\mbox{EQI}\left[\bx_c^{S+1}, \sigma^2(\bx_c^{S+1})\right]=\E_{Q^{S+1}}\left[\left(q^S(\bx_c^\star) - Q^{S+1}(\bx_c^{S+1})\right)^+\right]\,,
$$
where $(z)^+ = \max(0, z)$, $q^S(\bx_c)=m(\bx_c) + \Phi^{-1}(\beta)s(\bx_c)$ is the $\beta$-quantile from the current GP posterior~(\eqref{eq:gp-post}) with $\beta\in [0.5,1)$, $\bx_c^\star = \mathrm{arg min}_{\bx_c \in X_S} q^S(\bx_c)$ and $Q^{S+1}(\bx_c^{S+1})$ is the corresponding $\beta$-quantile when one additional observation $y(\bx_c^{S+1})$ is added to the data set. The choice of $\beta$ tunes the level of reliability wanted on the final result, setting $\beta=0.5$ means that the algorithm will compare design points based on the emulator's mean only, ignoring the uncertainty. Sequential design maximizing EQI is implemented in the \texttt{R} package \texttt{DiceOptim} \cite{picheny2014noisy}. 

It can be shown that the posterior for $Q^{S+1}(\bx_c^{S+1})$, conditional on $\by^S$ is also a GP \cite{picheny2013qbo} with posterior mean  
\begin{equation}\label{eq:quantile-mean}
m_{Q_{S+1}}\left(\bx_c^{S+1}, \sigma^2(\bx_c^{S+1})\right) = m(\bx_c^{S+1})+\Phi^{-1}(\beta)\sqrt{\frac{\sigma^2(\bx_c^{S+1})\times s^2(\bx_c^{S+1})}{s^2(\bx_c^{S+1}) + \sigma^2(\bx_c^{S+1})}}\,,
\end{equation}
and variance
\begin{equation}\label{eq:quantile-stdev}
s^2_{Q_{S+1}}\left(\bx^{S+1}, \sigma^2(\bx_c^{S+1})\right)=\frac{\left[s^2(\bx_c^{S+1})\right]^2}{s^2(\bx_c^{S+1})+ \sigma^2(\bx_c^{S+1})}\,,
\end{equation}
where $\Phi(\cdot)$ is the standard normal distribution function. Hence a closed form for EQI exists. Defining the improvement as $I = q^S(\bx_c^\star) - Q^{S+1}(\bx_c^{S+1})$,
\begin{align}
\mbox{EQI}\left(\bx_c^{S+1}\right) & = 
P(I> 0) \E_{Q^{S+1}}\left[q^S(\bx_c^\star) - Q^{S+1}(\bx_c^{S+1}) \mid I >0 \right] 
+ P(I\le 0) \E_{Q^{S+1}}\left[0 \mid I \le 0\right] \nonumber\\
& = \Phi_{Q_{S+1}}\left[q^S(\bx_c^\star) - E\left(Q^{S+1}(\bx_c^{S+1}) \mid I > 0\right)\right] \nonumber\\
& = \Phi_{Q_{S+1}} \left[q^S(\bx_c^\star) - \frac{m_{Q_{S+1}}\Phi_{Q_{S+1}} + s_{Q_{S+1}}\phi_{Q_{S+1}}}{\Phi_{Q_{S+1}}}\right]\nonumber\\
& = (q^S(\bx_c^\star) - m_{Q_{S+1}})
\Phi_{Q_{S+1}} + s_{Q_{S+1}}\phi_{Q_{S+1}}\,,\label{eq:EQI}
\end{align}
where
$$
\Phi_{Q_{S+1}} = \Phi\left(\frac{q^S(\bx_c^\star)-m_{Q_{S+1}}}{s_{Q_{S+1}}}\right)\,,
$$
and
$$
\phi_{Q_{S+1}} = \phi\left(\frac{q^S(\bx_c^\star)-m_{Q_{S+1}}}{s_{Q_{S+1}}}\right)\,,
$$
and $\phi(\cdot)$ is the probability function of the standard normal distribution. \eqref{eq:EQI} mimics the usual expression for EI.

Key to the EQI methodology is the `tunable precision' of the next observation, represented here by the variance $\sigma^2(\bx_c^{S+1})$. This is an effective way to allocate computing resources when faced with the common question of local/global search trade-off and inefficient sampling of iterative solvers \cite{AIJF2013}. Observations resulting from averaging a Monte Carlo sample were an original motivation for EQI \cite{picheny2013qbo} and lead to the important distinction between adding a design point at a new location and improving the Monte Carlo estimate by adding new simulations at an existing point via replication. This latter case can be handled by adjusting the variance of a subsequent data point at an existing location \cite{picheny2013abo} to be 
\begin{equation}\label{eq:re-sig}
\hat{\sigma}^2_{\mbox{rep}}(\bx_c) = \frac{\hat{\sigma}^2_{N}(\bx_c)\hat{\sigma}^2_{2N}(\bx_c)}{\hat{\sigma}^2_{N}(\bx_c) - \hat{\sigma}^2_{2N}(\bx_c)}\,,
\end{equation}
where, e.g., $\hat{\sigma}^2_{2N}(\bx_c)$ is obtained from~\eqref{eq:mcerror}.

When the Monte Carlo sample is from a collection of environmental variables, as in~\eqref{eq:mc_avg}, then $\sigma^2(\bx_c^{S+1})$ will typically be unknown until the sample has been generated. In this work, we set $\sigma^2(\bx_c^{S+1}) = \max_{\bx_c \in X_S} \sigma^2(\bx_{c})$ as a conservative choice that will result in lower EQI and favor exploitation (relative to a smaller choice of variance). A flowchart outlining the emulator-based optimization process for a single objective is given in Figure~\ref{opt:envir}.

\subsection{Multi-objective expected quantile improvement}
\label{subsec:Optimisation:multi_EQI}

For a non-trivial multi-objective optimization problem, usually no single solution exists that simultaneously optimizes each objective. A change from one solution to another that can improve at least one objective without making any other objectives worse is called a \emph{Pareto improvement}. A solution is defined as \emph{Pareto optimal} when no further Pareto improvements can be made \cite{hwang2012mod}. For conflicting objective functions, there exists a (possibly infinite) number of Pareto optimal solutions. These solutions are called \emph{non-dominated}, and the set of these solutions is a \emph{non-dominated set} or the \emph{Pareto optimal set}. The corresponding objective vectors define the \emph{Pareto front}.

Most approaches to GP-based multi-objective optimization build separate GP emulators for each of the $n$ objectives \cite{wagner2010expected}, with studies demonstrating little or no advantage in using emulators which assume dependence between objectives \cite{parr2013}. We follow this approach, and hence the joint posterior predictive density for the quantiles for each objective function at point $\bx$ is simply the product of the individual densities: 
\begin{equation}\label{eq:mvnorm}
\tilde{\phi}\left(q_1(\bx), \ldots, q_n(\bx)\right) = \prod_{i=1}^n\phi\left(\frac{q_i(\bx) - \mu_{Q_{S+1},i}}{s_{Q_{S+1}, i}}\right)\,,
\end{equation}
where $\mu_{Q_{S+1},i}, s_{Q_{S+1}, i}$ are the posterior mean and standard deviation, respectively, for the $\beta$-quantile from the $i$th objective function (see \eqref{eq:quantile-mean} and \eqref{eq:quantile-stdev}). 

\begin{figure}
\begin{subfigure}{.49\textwidth}
\centering
\scalebox{1.7}{\includegraphics[width=.5\textwidth
]{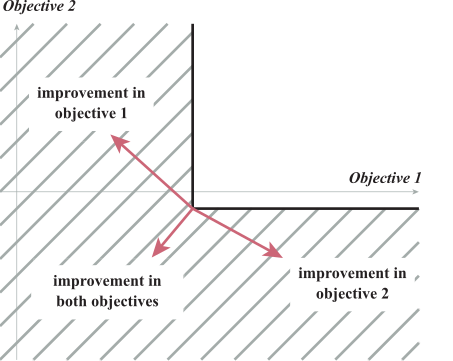}}
\caption{}
\label{fig:Dominating}
\end{subfigure}
\begin{subfigure}{.49\textwidth}
\centering
\scalebox{1.7}{\includegraphics[width=.5\textwidth
]{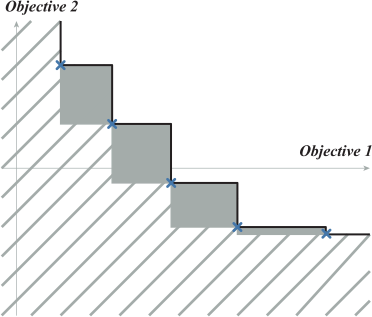}}
\caption{}
\label{fig:Pareto-set}
\end{subfigure}
\caption{(a) Improvements possible from a single point in the Pareto set. (b) A Pareto set of five non-dominated points (blue crosses) for a problem with two objectives. The solid line is the Pareto front. The shaded area shows where new points would augment the Pareto front, while the hatched area is where new points would dominate and replace the existing set of non-dominated points. Images adapted from Keane \cite{keane2006}}
\label{fig:int_by_parts}
\end{figure}

A variety of extensions and adaptations to EI for multi-objective optimization have been proposed \cite{zps2019}. Here, we focus on improving the Pareto front by finding solutions that dominate the current set (see Figure~\ref{fig:Pareto-set}) and extend EQI to multiple objectives by combining with the Euclidean distance-based method \cite{keane2006}. 

The Euclidean distance-based multi-objective EI criterion is well established for deterministic responses and has been applied successfully in a number of disciplines \cite{phtf2020,sobester2014engineering}. Other multi-objective EI formulations have also been proposed, notably, those based on hypervolume improvement \cite{egn2006}. These methods are `Pareto compliant', whereas Euclidean EI is not,  but may not have the same desirable Pareto-front space-filling characteristics as Euclidean EI. That is, Euclidean EI gives a higher expected improvement for a dominated point (so is not Pareto compliant) where there is a gap in the observed data, meaning the criterion will tend to populate these gaps with future observations \cite{wagner2010expected}. From a practical perspective, this Pareto-front space-filling characteristic is desirable in optimization.

Label the current $m$-point Pareto front $\mathcal{Q}_S = \{\bq_{1, S}^\star, \ldots, \bq_{m, S}^\star\}$, with $\bq_{j, S}^\star = (q^{S\star}_{j,1}, \ldots, q^{S\star}_{j,n})^{\sf T}$ the values of the $n$ objective functions for the $j$th quantile. Then the Multi-objective Euclidean EQI (MO-E-EQI) criterion chooses the next design point to maximize
\begin{equation}\label{eq:MOEEQI}
\mbox{MOEEQI}(\bx_c^{S+1}) = P_{S+1}(\bx_c^{S+1})\sqrt{(\bar{\boldsymbol{Q}}_{S+1}(\bx_c^{S+1}) - \tilde{\bq}_S)^{\sf T}(\bar{\boldsymbol{Q}}_{S+1}(\bx_c^{S+1}) - \tilde{\bq}_S)}\,, 
\end{equation}
where $P_{S+1}$ is the probability of improvement for $\bx_c^{S+1}$, obtained by integrating over the region that improves the Pareto front with respect to density in~\eqref{eq:mvnorm}, $\bar{\bQ}_{S+1}(\bx_c^{S+1})$ is the centroid of that integral and $\tilde{\bq}_S \in \mathcal{Q}_S$ is the closest Pareto point to the centroid.

For the common case of $n = 2$ objectives, the probability of improvement is available in closed form. \figref{fig:Dominating} illustrates the possible improvements from a single Pareto point, and \figref{fig:Pareto-set} demonstrates the region that requires integration. Evaluating the desired integral can be achieved by considering the various rectangles that comprise the hatched area in \figref{fig:Pareto-set}, moving from left to right and changing the limits of integration:
\begin{multline}
P_{S+1}(\bx_c^{S+1}) =
\Phi\left(\frac{q_{1,1}^{S*}-\mu_{Q_{S+1,1}}}{s_{Q_{S+1,1}}}\right) +\left\{1-\Phi\left(\frac{q_{m,1}^{S*}-\mu_{Q_{S+1,1}}}{s_{Q_{S+1,1}}}\right)\right\}
\Phi\left(\frac{q_{m,2}^{S*}-\mu_{Q_{S+1,2}}}{s_{Q_{S+1,2}}}\right)\nonumber\\
+\sum_{i=1}^{m-1}\Phi\left(\frac{q_{i+1,2}^{S*}-\mu_{Q_{S+1,2}}}{s_{Q_{S+1,2}}}\right)\left\{\Phi\left(\frac{q_{i+1,1}^{S*}-\mu_{Q_{S+1,1}}}{s_{Q_{S+1,1}}}\right)
-\Phi\left(\frac{q_{i,1}^{S*}-\mu_{Q_{S+1,1}}}{s_{Q_{S+1,1}}}\right)\right\}\,.\nonumber
\end{multline}
Similarly, when $n=2$ the elements of $\bar{\bQ}_{S+1}(\bx_c^{S+1})$ have the following form ($j=1,2$):
\begin{multline}\label{eq:centroid}
\bar{Q}_{S+1, j}(\bx_c^{S+1})P_{S+1}(\bx_c^{S+1}) = \int\limits_{-\infty}^{q_{1,1}^{S*}}\int\limits_{-\infty}^{\infty}q_j(\bx_c^{S+1})\boldsymbol{\phi}(q_1(\bx_c^{S+1}),q_2(\bx_c^{S+1}))\,\mathrm{d}q_2\mathrm{d}q_1\\
+ \sum\limits_{i=1}^{m-1}\int\limits_{q_{i,1}^{S*}}^{q_{i+1,1}^{S*}}\int\limits_{-\infty}^{q_{i+1,2}^{S*}}q_j(\bx_c^{S+1})\boldsymbol{\phi}(q_1(\bx_c^{S+1}),q_2(\bx_c^{S+1}))\,\mathrm{d}q_2\mathrm{d}q_1\\
+\int\limits_{q_{m,1}^{S*}}^{\infty}\int\limits_{-\infty}^{q_{m,2}^{S*}}q_j(\bx_c^{S+1})\boldsymbol{\phi}(q_1(\bx_c^{S+1}),q_2(bx_c^{S+1}))\,\mathrm{d}q_2\mathrm{d}q_1\,.
\end{multline}
A closed-form solution exists for~\eqref{eq:centroid} \cite{keane2006}. The required integration becomes increasingly intricate for $n>2$, and numerical approximations will suffer from a computational curse of dimensionality. In addition, the trade-offs required from decision makers for multi-dimensional problems become harder to conceptualize and require a higher cognitive burden; hence, two-dimensional problems are common in
many disciplines and represent an important class of problems.

Algorithm~\ref{alg:MO-E-EQI} outlines the sequential design process. Here, MO-E-EQI is calculated across a grid of control variables. This method is effective for the low dimensional problems considered here. For higher dimensional problems, some form of approximate numerical optimization would need to be employed, e.g., \cite{acebayes}. In Algorithm~\ref{alg:MO-E-EQI} we assume a fixed computational budget and run until this is exhausted. Alternatively, the algorithm could terminate when the MO-E-EQI objective function drops below a pre-determined threshold.

\begin{algorithm}
\caption{Multi-Objective Euclidean Expected Quantile Improvement (MO-E-EQI).}
\label{alg:MO-E-EQI}
\begin{minipage}{\textwidth}
\begin{algorithmic}

\STATE Construct initial design $X_S$ with $S$ initial design points in the control variables $\bx_c$ e.g., using a space-filling design.

\STATE Simulate $N$ samples from the joint distribution of the environmental variables, $\bx_e^1, \ldots, \bx_e^N$.

\STATE Run the computer model for each combination $\bx^j, \bx_e^r$ ($j = 1, \ldots, S;\,r = 1, \ldots, N$).
\FOR{$i=1,\ldots,n$}
    \FOR{$j=1,\ldots,S$}
        \STATE Return sample mean $y_i(\bx_c^j)$ (\eqref{eq:mc_avg}) and variance $\hat{\sigma}^2_{i, N}$ (\eqref{eq:mcerror}). 
    \ENDFOR
    \STATE Fit Gaussian Process emulator for $i$th response using data$\{y_i(\bx_c^j), \bx_c^j\}_{j=1}^S$ with weights $\{\hat{\sigma}^2_{i, N}(\bx_c^j)\}_{j=1}^S$.
\ENDFOR
\FOR{$q =1,\ldots,N_{iter}$}
    \STATE Find the Pareto front based on the current emulator-based quantiles for all observations.
    \STATE Calculate MO-E-EQI \eqref{eq:MOEEQI} at a grid of control variable locations.
    \STATE Select the point $\bx_c^\dagger$ with highest MO-E-EQI.
    \STATE Simulate $N$ samples from the joint distribution of the environmental variables, $\bx_e^1, \ldots, \bx_e^N$.
    \STATE Run the computer model at each design location determined by the combinations of $\bx_c^\dagger$ with sample values of the environmental variables.
    \FOR{$i=1,\ldots,n$}
        \STATE 
        Return sample mean $y_{i}(\bx_c^\dagger) = \frac{1}{N} \sum\limits_{r=1}^{N}h_i(\bx_c^\dagger,\bx_e^r)\text{ and sample variance } \hat{\sigma}_{i, N}^2(\bx_c^\dagger)$
    \ENDFOR
    \IF{$\bx_c^\dagger$ is a new design point}
        \STATE Add it to the current design $X_{S+q}=X_{S+q-1} \cup \bx_c^\dagger$.
        \STATE Update Gaussian Process emulator with data $\{y_i(\bx_c^\dagger),\bx_c^\dagger\}$ and weight $\hat{\sigma}_{i, N}^2(\bx_c^\dagger)$.
    \ELSE
        \STATE Append the repeated design point with mean $y_{i}(\bx_c^\dagger)$ and variance from \eqref{eq:re-sig}. 
        \STATE Update Gaussian Process emulator with the updated observation and variance values.
    \ENDIF
\ENDFOR
\RETURN  The Pareto front based on the current emulator-based quantiles for all observations and the final design $X_{S+N_{iter}}$ that corresponds to the Pareto front points. 
\end{algorithmic}
\end{minipage}
\end{algorithm}

This algorithm is deliberately quite ``aggressive" and only searches for points that will \emph{dominate} at least one of the current solutions. However, if the integration limits in~\figref{fig:Pareto-set} are changed to include the shaded rectangles (that are currently excluded from the limits of integration), a ``non-aggressive" algorithm will result in filling in gaps between current non-dominated solutions. Using this algorithm, there will generally be more points on the Pareto front. We focus here on the results of only using the aggressive algorithm, and demonstrate how the algorithm effectively gets close to the true Pareto front in very few steps. However, a combination of the two approaches might be desirable depending on the goal. Such an approach would include several steps of the aggressive algorithm initially to identify the true Pareto front quickly, followed by some steps of the non-aggressive algorithm to fill in more points on the Pareto front. Some exemplar results from applying such a combination to the example in Section~\ref{sec:Optimsation:illustrative_example} can be found in Appendix~\ref{A:non_aggressive}.

\section{Illustrative example}
\label{sec:Optimsation:illustrative_example}

We use the following functions to illustrate our methodology on a two-objective optimization problem with two control and two environmental variables:
\begin{align}\label{eq:toy}
\begin{split}
    h_1(\bx_c,\bx_e)&= 1-\sin(x_{c_1})+a\cdot\cos(x_{e_1})+\frac{x_{c_2}+x_{e_2}}{10}\\
    h_2(\bx_c,\bx_e)&= 1-\cos(x_{c_1})+a\cdot\sin(x_{e_1})+\frac{x_{c_2}+x_{e_2}}{3},
\end{split}
\end{align}
where $\bx_e^{\sf T}=(x_{e_1},x_{e_2})$, $x_{e_1}\sim U(-\pi, \pi)$ and $x_{e_2}\sim \mathcal{N}(0, 0.5^2)$, $\bx_c^{\sf T}=(x_{c_1},x_{c_2})$, $x_{c_1}\in[0,\frac{\pi}{2}]$ and $x_{c_2}\in[0,1]$, and parameter $a\ge 0$. In this example, the output noise $\sigma^2(\bx_c)$ is constant for all $\bx_c$ (stationary) due to the lack of interactions between the control and environmental variables. These functions were chosen to illustrate the impact of variables with both nonlinear ($x_{c_1}, x_{e_1}$) and linear ($x_{c_2}, x_{e_2}$) effects, and to assess the robustness of MO-E-EQI to different levels and structure of uncertainty through the choice of $a$. We present the response contours in \figref{fig:ground_truth} in Appendix~\ref{app:truth} with environmental variables' noise integrated out. This could be considered the `ground truth'.

\begin{figure}
\centering
\begin{center}
\includegraphics[width=\textwidth]{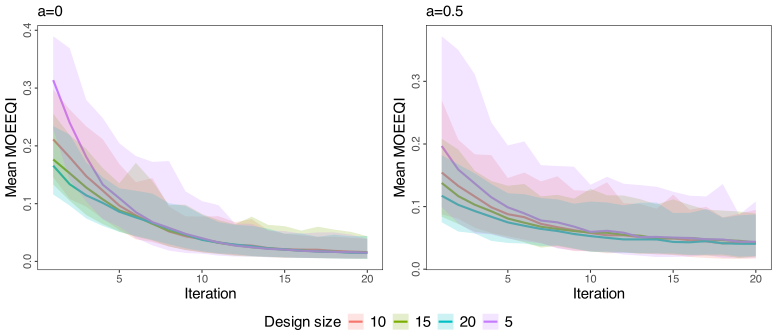}
\end{center}
\caption{Evolution of the MOEEQI metric over 20 iterations for different initial design sizes for $a=0$ (left) and $a=0.5$ (right). The solid lines are the mean performance over 100 repetitions and shaded areas are 90\% confidence intervals.}
\label{fig:EI_res}
\end{figure}

GP emulators~\ref{eq:gp-post} are built using Monte Carlo estimates of $f_i(\bx_c^l) = \E(h_i(\bx_c^l,\bx_e))$, for $i=1,2$; $l=1,\ldots,S$,  with $N=10$ simulations from $\bx_e$ (using function \texttt{km} from \texttt{R} \cite{R} package \texttt{DiceKriging} \cite{rgd2012}). The initial design was an $S = 5$ maximum projection Latin hypercube \cite{jgb2015} (from the \texttt{MaxPro} \cite{bj2018} \texttt{R} package). The use of a modestly sized initial design reflects the typical use case of (multi-objective) Bayesian optimization to correctly and efficiently identify the true optimum or Pareto front, rather than build globally effective emulators for functions $f_1$ and $f_2$. We explore the impact of the choice of the size of the initial design later in \figref{fig:EI_res}.

For each iteration of the MO-E-EQI loop, objective function in~\eqref{eq:MOEEQI} is evaluated on a two-dimensional grid of $\bx_c$-values, with 100 points in each dimension, to find the next design point $\bx_c^{S+1}$ that maxmizes~\eqref{eq:MOEEQI}. Monte Carlo estimates of $f_1(\bx_c^{S+1})$ and $f_2(\bx_c^{S+1})$ are then obtained and added to the vectors of results $\by^{S+1}_1$ and $\by^{S+1}_2$.

\begin{figure}
\centering
\begin{subfigure}{.5\textwidth}
\begin{center}
\includegraphics[width=\textwidth]{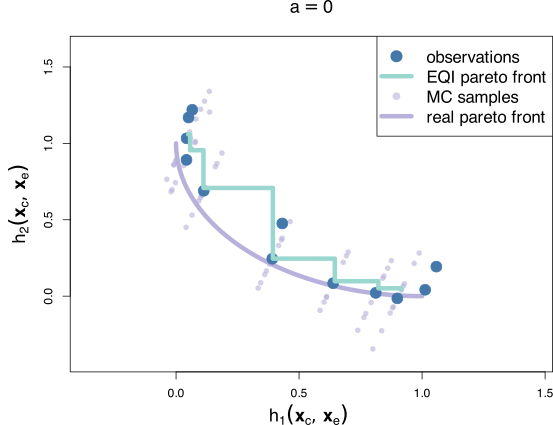}
\end{center}
\caption{}
\end{subfigure}%
\begin{subfigure}{.5\textwidth}
\begin{center}
\includegraphics[width=\textwidth]{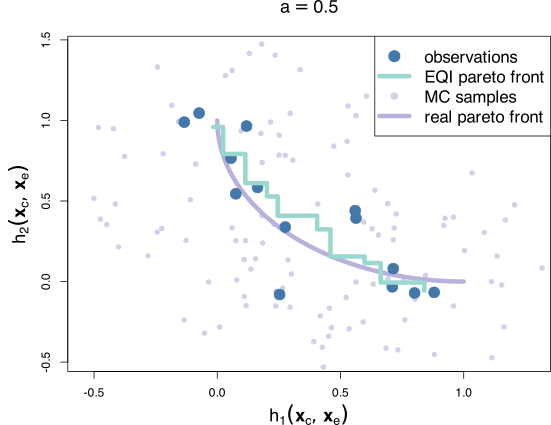}
\end{center}
\caption{}
\end{subfigure}%
\caption{MO-E-EQI results for 9 iterations (plus 5 initial runs) for two test functions $h_1$ and $h_2$ from \eqref{eq:toy} and two values of the tuning parameter $a$: (a) $a=0$ and (b) $a=0.5$.
The plots show all $S=14$ observations, the real Pareto front, the current quantile-based Pareto front and all the samples from the joint distribution of $h_1$ and $h_2$ for each of the $S=14$ design points.}
\label{test_function}
\end{figure}

The results of running the algorithm assuming budget for $N_{iter} = 9$ additional observations, with $\beta=0.7$ for each of two different values of $a=0$ and $0.5$, are shown in \figref{test_function}. The estimated Pareto fronts in these plots are the combined $\beta$-quantile values $q^S_{i}(\bx_c)=m_{i}(\bx_c) + \Phi^{-1}(\beta)s_{i}(\bx_c)$ ($i=1,2$) using the final emulator means and variances with $S=14$ observations. Note that for noisy observations such as these, observations identified as belonging to the Pareto set will not necessarily lie on the Pareto front calculated from the quantiles. Also shown are the true Pareto fronts, the $S=14$ observations and samples from the joint distribution of $h_1$ and $h_2$ for all $S=14$ observations.

\begin{figure}
\centering
\begin{subfigure}{.5\textwidth}
\begin{center}
\includegraphics[width=\textwidth]{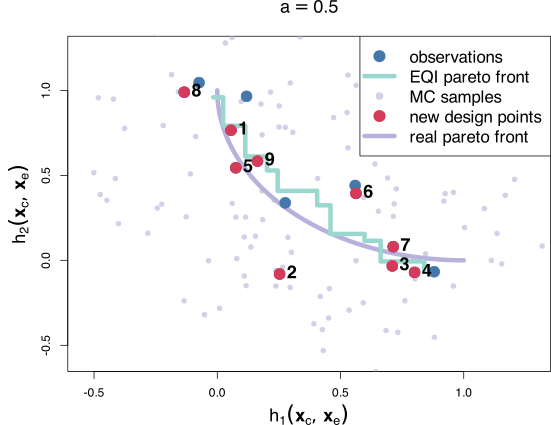}
\end{center}
\caption{}
\end{subfigure}%
\begin{subfigure}{.5\textwidth}
\begin{center}
\includegraphics[width=\textwidth]{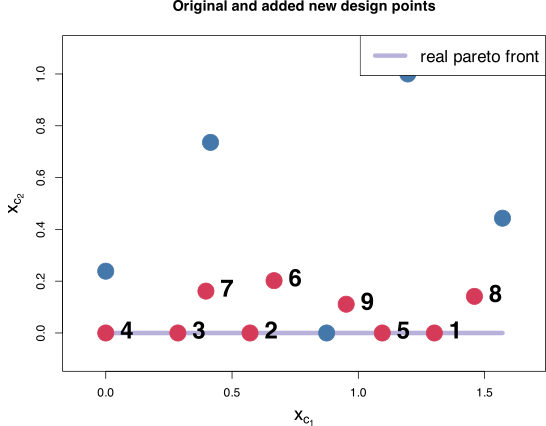}
\end{center}
\caption{}
\end{subfigure}%
\caption{Sequential design points identified by the MO-E-EQI criterion. The original design had 5 points and 9 more points were added sequentially by maximizing MO-E-EQI. Panel (a) demonstrates the sequentially added 9 points along with the original 5 points, pooled observations, the real Pareto front, the current quantile-based Pareto front and all the samples from the joint distribution on $h_1$ and $h_2$ for each of the $S=14$ design points; and panel (b) shows the addition of new points in the control variable input space.}
\label{seq:design}
\end{figure}

\figref{seq:design} shows the sequence of points added via maximization of MO-E-EQI, and their locations in the $x_{c_1}-x_{c_2}$ space. The majority of the points on the estimated Pareto front are added when $x_{c_2}$ is zero or small (due to the linear effect of this variable in both functions), and for a variety of values of $x_{c_1}$ (due to the competing sine and cosine functions in $h_1$ and $h_2$, respectively). 

The Pareto fronts are quite well identified for both values of the parameter $a$, although predictably the results are a little worse when there is larger and more complex noise ($a=0.5$). When $a=0$, there is a considerable structure in the joint distribution of the random functions $f_1$ and $f_2$ arising from the shared environmental variable $x_{e_2}$. When there is greater variability via a nonlinear function, as is the case when $a=0.5$, this structure disappears but the Pareto front is still reasonably well identified in a small number of runs. In common with many Bayesian optimization applications, these results were achieved with an initial design of only modest size ($S = 5$). In fact, increasing the initial design size, using a larger MaxPro LHS, does not materially change the results (see Figure~\ref{fig:EI_res}). From this figure we also see that, for this example, the value of objective function~\eqref{eq:MOEEQI} for MO-E-EQI stabilizes by $N_{iter}=15$ iterations for both $a=0$ and $a=0.5$.

\begin{figure}
\centering
\begin{center}
\includegraphics[width=\textwidth]{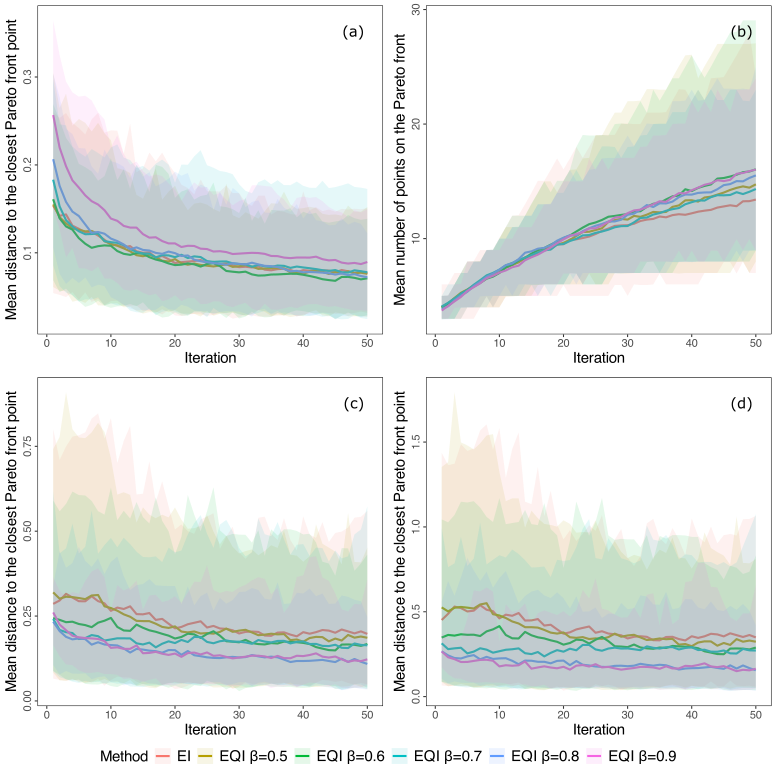}
\end{center}
\caption{(a) Mean distance from identified points to the nearest true Pareto front point. (b) Mean number of points on the Pareto front.
(c)-(d) Mean distance from identified points to the nearest true Pareto front point where the distances that overestimate the Pareto front are penalized by a factor of 5 (c) and a factor of 10 (d).}
\label{fig:EI_dist}
\end{figure}

As a benchmark for the MO-E-EQI method, we compare a simple extension of ``plug-in'' expected improvement \cite{picheny2013abo} to the multi-objective setting (MO-E-EI) using the example when $a=0.5$. Here, the Pareto front is estimated using the current GP mean and the next design point chosen to minimize the expected Euclidean distance to this current front. The GP prior is updated assuming noisy observations, as in~\eqref{eq:gp-post} but future noise is ignored, $\sigma^2(\bx_c^{S+1}) = 0$. Results are shown in Figure~\ref{fig:EI_dist} where MO-E-EQI for various $\beta$-quantiles is compared to MO-E-EI in terms of the mean distance to the nearest Pareto point and the mean number of points on the Pareto front from 100 applications of each method, adding an additional $N_{iter}=50$ sequential points. From Figures~\ref{fig:EI_dist}(a) we see that MO-E-EI is competitive with EQI-based criteria in terms of the decrease in mean distance to the front. However, we see from Figure~\ref{fig:EI_dist}(b) that the mean number of points on the Pareto front is fewer for MO-E-EI, as it is a less conservative method and tends to explore further from the Pareto front. In general, while the distance to the Pareto front stabilizes after about 40 iterations for all criteria, the number of points on the front continues to increase. 

We can further explore the differences between criteria by assessing whether the estimated Pareto front is under- or over-estimated. In Figures~\ref{fig:EI_dist}(c) and~(d), we penalized distances from the Pareto front that overestimated the front by a factor of either 5 or 10 respectively. The differences between more and less conservative methods is clear, with the aggressive MO-E-EI and MO-E-EQI with $\beta = 0.5$ strategies tending to overestimating the front and hence having larger penalized distances than the conservative MO-E-EQI methods with higher $\beta$ values. MO-E-EQI with $\beta = 0.9$ is the most conservative approach, favoring exploitation over exploration and hence choosing more repeated design points. Clearly, differences between methods will be example specific, as has also been noted in the single objective case \cite{picheny2013abo}.

\section{Anthrax release model example}
\label{sec:Optimisation:anthrax}
Bacillus anthracis is a spore-forming bacterium that can cause a variety of life-threatening infections in humans and other mammals. Anthrax is spread by contact with the bacterium's spores. A large number of fatalities could result from a targeted release near a major city due to the very small number of spores required to cause inhalation anthrax (hundreds to thousands). Inhalation anthrax is particularly deadly, with historical mortality rates over 85\% \cite{kortepeter2001usamriid,hendricks2014centers} and $\sim$45\% mortality when treated early \cite{holty2006systematic,hendricks2014centers}. Often, many fatalities from inhalation anthrax are when the first stage is mistaken for the cold or flu and the victim does not seek treatment until the second stage, which is 90\% fatal. A recommended intervention is oral antibiotics for known or imminent exposure \cite{kortepeter2001usamriid}. 

The goal of an effective intervention after an anthrax attack is to prevent illness when possible by treating exposed and potentially infected people when still curable (i.e.,~it is important to distribute antibiotics before toxin levels could rise sufficiently to cause sepsis and death). In order to optimize a potential intervention, we would need to model the dispersion first, then the epidemiology and then how a potential intervention changes the epidemiology. Additionally, to define a sensible intervention scenario, we need to calculate the cost of an intervention and the limitations in terms of resources (e.g.,~people or drugs). 

\subsection{Dispersion-dose-response-intervention simulator}
A variety of mathematical models have been developed to describe and predict both the dispersion of anthrax spores and the dose-response on an affected population \cite{wilkening2006sverdlovsk, brookmeyer2005modelling,wein2003emergency, reshetin2003simulation}. In addition, many models can incorporate the impact of interventions, such as the preventative use of antibiotics. We apply the MO-E-EQI methodology to a particular anthrax modeling system to find a Pareto front trading off fatalities against resource available for intervention.  

The anthrax release model used here simulates a dispersion of the anthrax spores over space and models dose response (the probability of becoming ill, given the dose received) in the population. In addition, it can model an intervention: given certain resources some of the people that would die without an intervention survive. We are interested in optimizing a potential course of action (intervention) in case of an anthrax release. 

Interventions applied to treat an anthrax outbreak have an associated opportunity cost in terms of their impact on other aspects of the healthcare system, e.g., clinical staff being diverted to administer antibiotics. We define an abstract notion of cost in our intervention model (Section~\ref{subsec:Optimisation:cost_model}) to comparatively measure different allocations of resources to the anthrax response. 

\subsubsection{Dispersion model}
\begin{figure}
\centering
\begin{subfigure}{.6\textwidth}
\begin{center}
\includegraphics[width=1\textwidth]{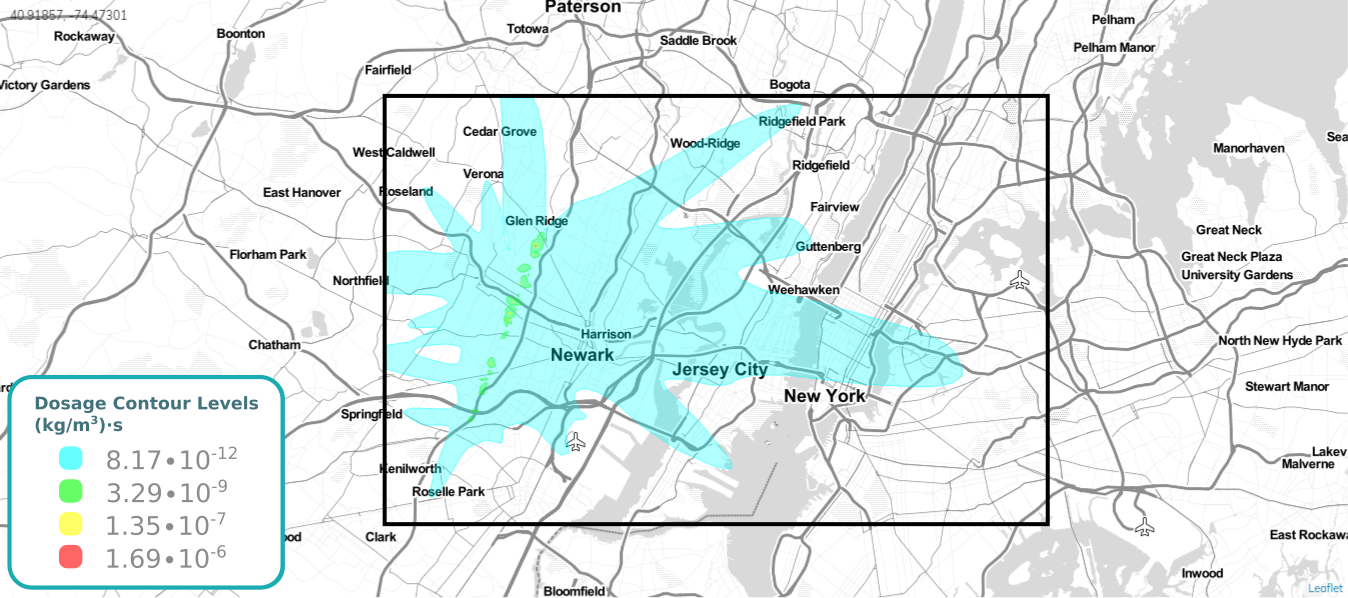}
\end{center}
\caption{}
\label{UDM_plume}
\end{subfigure}%
\begin{subfigure}{.4\textwidth}
\begin{center}
\includegraphics[width=.9\textwidth]{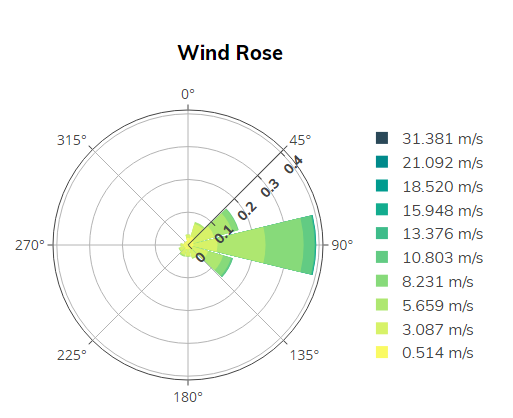}
\vspace{.25cm}
\end{center}
\caption{}
\label{wind_rose}
\end{subfigure}%
\caption{Left panel: example plumes of dosages resulting from running the dispersion model for a sample of values of the environmental variables defined in Table~\ref{tab:model_param}. The black box bounds the modeling area, covering approximately 120 km$^2$. Right panel: This wind rose is a visual representation of the wind speed and wind direction distributions.}
\end{figure}

In this hypothetical scenario, the spores spread in space, creating a plume that will represent dosages at each point in space. The shape of such a plume will depend on features including wind speed, wind direction and source mass; see~\figref{UDM_plume} for an example and Table~\ref{tab:model_param} for a list of input parameters. The studied anthrax model uses a Gaussian puff model \cite{llewelyn1983, OOYMMSI2001} implemented by the UK Defence Science and Technology Laboratory. For illustration, we assume terrain and population spread taken from an area of New Jersey around Newark International Airport; see Figure~\ref{UDM_plume}. The inputs to the dispersion model are regarded as environmental variables (Table~\ref{tab:model_param}) and impact the total number of people affected and the dosages that they receive. Those dosages are passed into the dose-response model to predict how the individuals in the population will be affected.  

\subsubsection{Dose response model} 
In the case of inhalation anthrax infection, clinical illness can develop any time between 2 and 40 days after exposure \cite{wilkening2008mti} with the average times ranging between 7-12 days. A log-normal distribution defines the incubation period with dose-dependent $\mu$ and $\sigma$. After dose response is calculated, the population is split into those who will develop symptoms ($E$), and those who will not ($G$). For those who develop symptoms, we sample from a distribution governing time from symptom onset to hospitalization and time from hospitalization to death \cite{holty2006systematic}. 

\subsubsection{Interventions}\label{subsec:Optimisation:cost_model}
Intervention times are simulated from a uniform distribution (between intervention start day and intervention end day) and compared to the symptom onset time. For illustration, an intervention that occurs before the symptomatic period for an individual is assumed to be 100\% efficacious and instantaneous, and prevents an individual from becoming symptomatic. Uncertain efficacy could be incorporated as another environmental variable. The model outputs the number of fatalities, given an intervention start and end days. Intervention start day is a controllable variable, and intervention end day is a function of two other controllable variables: $X$, the number of antibiotic collection centers (ACC) and $S_{\text{ACC}}$, the staff per ACC (Table~\ref{tab:model_param}).

To provide a second, competing, objective to fatalities, we additionally model the intervention cost, in the spirit of a health economic analysis. The purchase cost of the medical countermeasure stockpile, the resulting storage costs and the cost of administering the drugs in an event can be evaluated. 
      
We assume an affected population of size $P$, with each individual requiring $\nu$ doses of the medical countermeasure. It is also assumed that a percentage of the countermeasures will be wasted, with wastage multiplier $\rho$, so that total doses for complete coverage of the affected population would be $\nu\rho P$. 
        
This means for purchase cost $a$ per dose the purchase cost is $c_p=a\nu\rho P$ per procurement. Assuming a planning horizon for procurement of $Y$ years and a countermeasure shelf life of $T$ years then on average the total purchase cost is $(c_p Y)/T$. Therefore total purchase costs over a horizon of $Y$ years are $C_p=a\nu\rho PY/T$. We assume an outbreak has no impact on re-procurement costs.
      
If an outbreak occurs in the time frame then the countermeasure supply will have to be administered, the average total administration cost is $C_A$. This covers countermeasure collection center set up costs, $g$, (delivering, resupplying, hire of venue and staff training) that are incurred per center and staff administration costs, $d$, incurred per meeting with qualified staff per countermeasure delivered. If we assume there are $X$ collection centers and $S$ staff then $C_A=(gX+dS\nu P) Y$. The staff $S$ is the total number and might be broken down by staff per collection center if necessary.

Hence, the total cost (ignoring discounting) is then 
\begin{align}\label{eq:cost}
C & = C_A + C_P\\\nonumber
& = \left(\left(\frac{gX}{\nu P} + dS \right) + \frac{a\rho}{T}\right)\nu PY\,.
\end{align}

\subsection{Multi-objective optimization with uncertain inputs}
\label{subsec:Optimisation:example_EQI}

We employ the MO-E-EQI methodology to find optimal intervention settings for the anthrax modeling problem. As discussed above, and summarized in Table~\ref{tab:model_param}, the dispersion model inputs are treated as environmental variables and the intervention parameters as control variables. Table~\ref{tab:model_param} also specifies the sampling distributions for the environmental variables. The wind rose is discretized into a two-way table of frequencies of directions and speeds. A direction is sampled from the density approximated from the table margin, and then a speed is sampled from the corresponding table row according to the normalized frequencies. 

The two, competing, modeling outcomes of interest are (i) the number of fatalities and (ii) the intervention cost. We fixed the intervention start day because there is no cost associated with it in our model and consequently the algorithm was always choosing the earliest day possible.

If an optimisation problem in~\eqref{opt_prob} has upper bound constraints, they are usually presented as an additional inequality:
\begin{equation}\label{}
\begin{array}{rll}
\min & f_i(\bx_c) & \mbox{for } i=1,\ldots, n\\
\text{subject to: }
& \mbox{lb}_k\leq x_{c,k} \leq \mbox{ub}_k & \mbox{for } k=1,\ldots,v,\\
\text{and: } & f_i(\bx_c)\leq b_i.
\end{array}
\end{equation}

When finding the solutions on the current Pareto front in the noiseless case, the algorithm discards all the solutions that are outside of the constraints. When working with noisy observations, all potential solutions outside of the constraints adjusted for noise are similarly discarded. This means that a potential solution is discarded if its quantile is outside of the constraint adjusted for noise: $m_{Q_{S+1},i}\left(\bx_c^{S+1},\sigma^2(\bx_c^{S+1})\right)\geq b_i+\Phi^{-1}(\beta)
s^2_{Q_{S+1}}\left(\bx^{S+1}, \sigma^2(\bx_c^{S+1})\right).$
In our problem, the fatalities and intervention cost objectives were constrained to not exceed the set values of $10^5$ for deaths and $6\times 10^8$ for cost. 

Analogous to the example in \Secref{sec:Optimsation:illustrative_example}, we will use Monte Carlo sample means as observations. The variance of a single observation for both objectives was estimated during the initial construction of the emulator when the model was run at each design location with uncertain inputs. The maximum estimated sample variances were chosen and the baseline standard deviations were calculated at $\hat{\sigma}_{1, 1}(x_c) \approx2.949\cdot10^3$ and $\hat{\sigma}_{2, 1}(x_c)\approx4.996\cdot10^8$ using samples from $\bx_e$. Those standard deviations are a representation of the uncertainty in the dose-response model's outputs, given uncertain inputs to the dispersion model. In this example, at each step of the algorithm, the same fixed number of model runs was performed with $N=50$ sampled values of $\bx_e$, so the corresponding variances reduce to $\hat{\sigma}^2_{1, N}(x_c) \approx\frac{(2.949\cdot10^3)^2}{50}$ and $\hat{\sigma}^2_{2, N}(x_c) \approx\frac{(4.996\cdot10^8)^2}{50}$. The number of samples is chosen primarily for illustration. Clearly, $N>50$ would result in smaller variances; variances at some combination of control variables will be reduced by replication.  

\begin{table}
\begin{center}
\begin{tabular}{lll}
\multicolumn{3}{l}{Environmental variables}\\
\hline
 & Distribution & Units \\ 
 \hline
Release mass & $U[ 0.001 , 0.008]$ & kg \\
Wind direction & Wind rose (\figref{wind_rose}) & degrees \\         
Wind speed & Wind rose (\figref{wind_rose}) & m/s \\
Release location (longitude) & $U[-74.25 , -74.2]$ & degrees \\
Release location (latitude) & $U[40.7 , 40.8]$ & degrees \\
Release duration & $U[0,10]$ & seconds \\
Dry particulate matter fraction & $U[0,1]$ & -- \\
\hline \\
\multicolumn{3}{l}{Control variables}\\
\hline
 & Range & Units \\
\hline
Number of ACC ($X$) & $[1,10]$ & -- \\
Staff per ACC ($S_{\text{ACC}}$) & $[5,40]$ & -- \\
\hline
\end{tabular}
\end{center}
\caption{Environmental and control variables for the anthrax example.}
\label{tab:model_param}
\end{table}

\begin{figure}[ht]
\centering
\begin{subfigure}{.5\textwidth}
\begin{center}
\includegraphics[width=\textwidth]{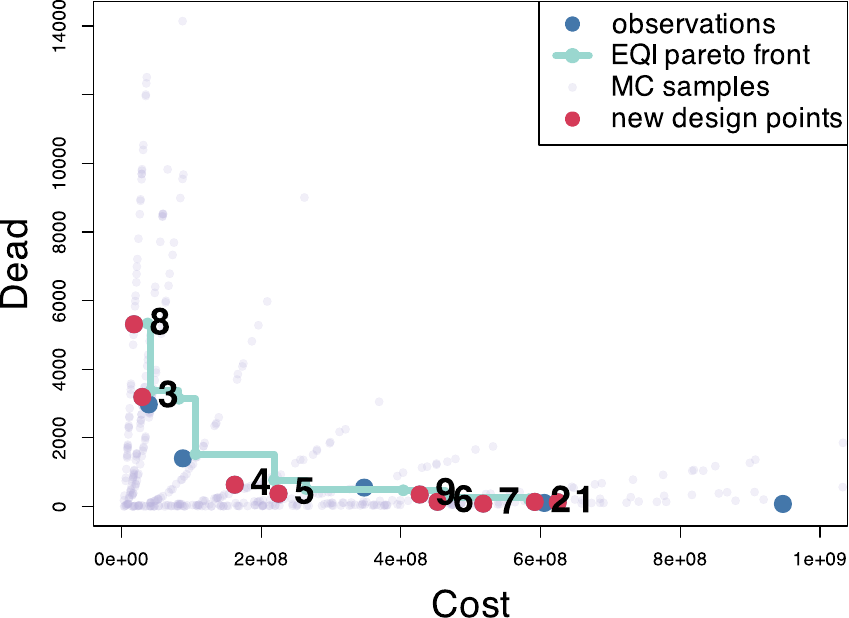}
\end{center}
\caption{}
\end{subfigure}%
\begin{subfigure}{.5\textwidth}
\begin{center}
\includegraphics[width=\textwidth]{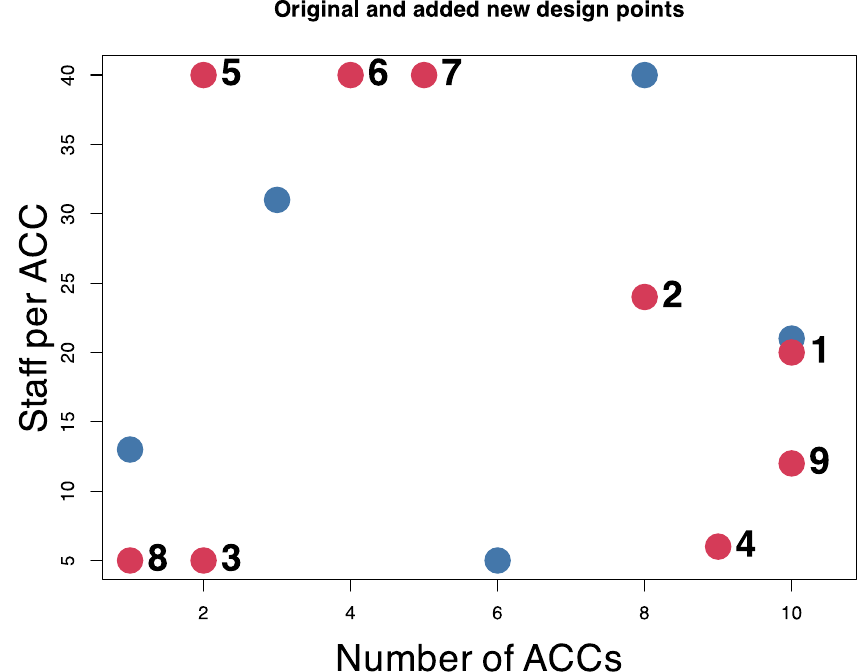}
\end{center}
\caption{}
\end{subfigure}%
\caption{Results of applying the EQI algorithm to the anthrax model with uncertain inputs. The two objectives for optimization are the total number of fatalities and the cost of intervention. Panel (a) shows the original five observations in blue and all nine added points with the order of addition (red points with corresponding numbers). The current Pareto front identified by the algorithm is presented in green and the Monte Carlo Samples corresponding to each point on the Pareto front in presented in purple; and panel (b) shows the
addition of new points in the control variable input space}
\label{Dead_vs_Benefit}
\end{figure}

Results from applying MO-E-EQI with $\beta=0.8$ (to promote conservative solutions with less uncertainty) are presented in \figref{Dead_vs_Benefit}. The original design was a two-dimensional Latin hypercube design with maximum projection space filling \cite{jgb2015} of size $S = 5$ in the space of $(X,S_{\text{ACC}})$. An additional $N_{iter} = 9$ points were added using MO-E-EQI. If the maximum expected improvement occurred by repeating a combination of control variables, the variance at that point is updated via~\eqref{eq:re-sig}. 

\figref{Dead_vs_Benefit} shows all 14 observations. The quantile values that were identified as part of the final Pareto front are connected by a step function. Additionally, the Monte Carlo samples  that correspond to the points on the final Pareto front are presented. It is clear that the uncertain inputs have a major effect on the model predictions. However, the algorithm addresses this uncertainty and is still able to find an optimal front using the average model predictions.

It is clear that the uncertainty in the two outputs is not constant across the Pareto front. In particular, there is greater variability in the intervention cost for higher values of this output and lower variability in the number of fatalities. Uncertainty in the intervention cost for these resource levels is driven by variation in the size of the affected population, which in turn is determined by the uncertain environmental input variables. When less resource is being allocated to interventions, there is much less uncertainty in cost (it is essentially capped regardless of the numbers of affected people) but consequently a much greater range of fatalities are possible, again depending on the uncertain values of the environmental variables. See \cite{chapman_a2014, chapman_b2014} for alternative plots to assess uncertainty in Pareto solutions and their role in decision making.

\section{Discussion and future research}
\label{sec:Optimisation:Summary}

Many policy decisions in the health, environment and government arenas are increasingly relying on (i) complex mathematical modeling via computer codes and (ii) the identification of optimal decisions in the presence of noise. We have presented and applied an approach to optimal decision making that overcomes issues around computational expense of evaluating these models and incorporates the uncertainty caused by uncontrolled environmental variables. A Pareto front approach allows subject matter experts to directly understand trade-offs between competing objectives in a clear and (relatively) unambiguous manner and hence is arguably more practically useful than compound, constrained or robust optimization \cite{gorissen2015practical,ben2009robust} that requires interpretation of abstract tuning parameters. This was certainly our experience in discussing the hypothetical anthrax example with public health experts. If experts can provide a weighting, or weightings, of individual objectives, these can be incorporated post-hoc by assessing weighted combinations of objective function values both numerically and graphically \cite{lu2014}. Such weights could also be incorporated into the optimization algorithm \cite{lyzz2024}, for example, by obtaining the probability of improvement $P_{S+1}(\boldsymbol{x}_c^{S+1})$ via integration with respect to a measure reflecting these weights. Such methodology is a possible route for future research; however we believe that for many applications post-hoc consideration of weightings may be more appropriate, using the whole Pareto front unbiased by possibly uncertain priorities.

There are a number of other areas of future research, including extensions of other multi-objective improvement methods, e.g., hypervolume improvement, to incorporate noise from environmental variables \cite{ho2022}; more sophisticated incorporation of non-constant variance, e.g., using heterogeneous Gaussian process models; methods for higher dimensional responses; different covariance functions to address responses of varying smoothness; and robust optimization approaches, e.g., accounting for Monte Carlo mean and variance estimates, for multi-objective problems. 

\begin{figure}[ht]
\centering
\includegraphics[width=\textwidth]{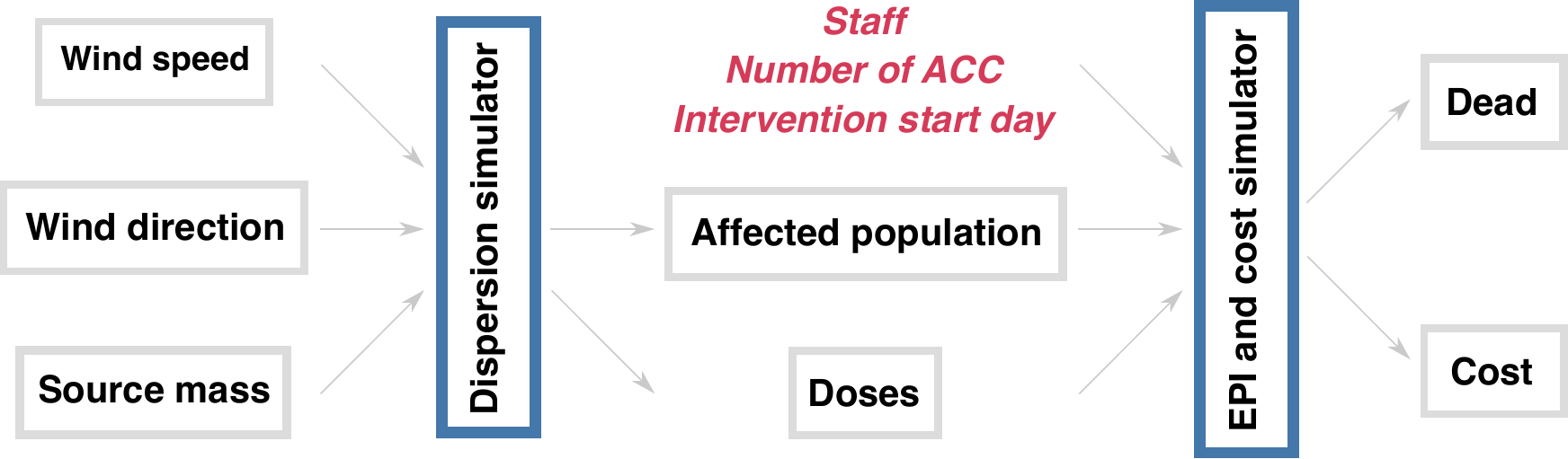}
\caption{The schematic of the two anthrax models. In gray boxes are the uncertain inputs and outputs for each model. In red are the intervention parameters which we wish to optimize.}
\label{chain_opt}
\end{figure}

A feature of the anthrax example is the composition of the modeling system from a number of individual, but linked, simulators; in this case, a dispersion and a dose-response model. The use of such a building block approach is common in modeling complex systems \cite{kbw2018, JW2021}. More details about the chain of simulators in the anthrax case is given in Figure~\ref{chain_opt}.

One obvious implication of such a modeling chain is the possibility of constructing and utilizing separate emulators for each simulator. Such a strategy may be particularly advantageous when there is disparity in the computational costs of running the different simulators in the chain. For the anthrax example, the dispersion simulator is substantially more expensive to run. If a reliable emulator can be constructed for the dispersion simulator, the predictive distribution from this emulator can be used as inputs to the Dose Response model (or its emulator). Such an approach to reducing computational cost would come with greater uncertainty, due to the use of emulated inputs to the dose-response model but the impact could be tuned via the choice of design \cite{MG2021} and sample size.

\section*{Acknowledgments}
We are grateful to CrystalCast project members for invaluable discussions, comments, and provision of the simulators for the dispersion dose-response application. Particular thanks are due to Professor Veronica Bowman and Dr Hannah Williams (Defence Science and Technology Laboratory, UK), Dr Daniel Silk, Dr Samuel Jackson (University of Durham, UK), Dr Ian Hall (University of Manchester, UK), Dr Thomas Finnie (UK Health Security Agency), and Sacha Darwin and Helen Adams (Riskaware).

\bibliographystyle{unsrt}  
\bibliography{opt_bib}

\appendix
\section{Combining the aggressive and non-aggressive algorithms}\label{A:non_aggressive}

\begin{figure}[!ht]
\centering
\begin{subfigure}{.5\textwidth}
\begin{center}
\includegraphics[width=\textwidth]{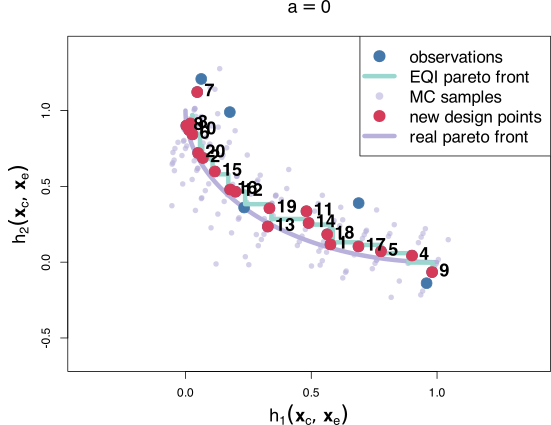}
\end{center}
\caption{}
\end{subfigure}%
\begin{subfigure}{.5\textwidth}
\begin{center}
\includegraphics[width=\textwidth]{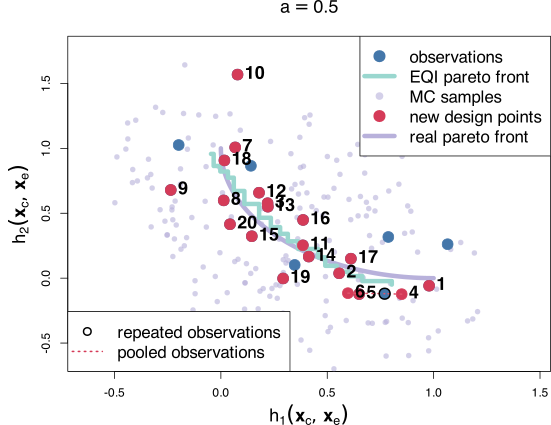}
\end{center}
\caption{}
\end{subfigure}%

\begin{subfigure}{.5\textwidth}
\begin{center}
\includegraphics[width=\textwidth]{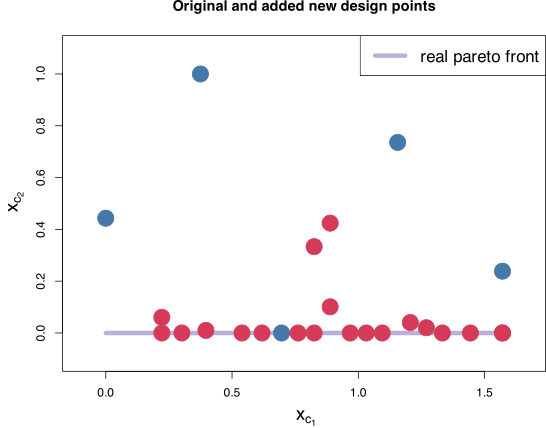}
\end{center}
\caption{}
\end{subfigure}%
\begin{subfigure}{.5\textwidth}
\begin{center}
\includegraphics[width=\textwidth]{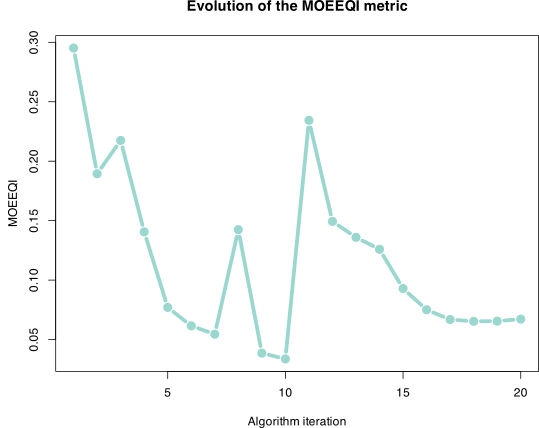}
\end{center}
\caption{}
\label{MOEEQI_evolution}
\end{subfigure}%
\caption{
Sequential design points identified by the MO-E-EQI criterion. The original design had 5 points and 20 more points were added sequentially by maximizing MO-E-EQI (10 iterations of aggressive and 10 non-aggressive approach). Panels (a) and (b) demonstrate the sequentially added 20 points along with the original 5 points, pooled observations, the real Pareto front, the current quantile-based Pareto front and all the samples from the joint distribution on $h_1$ and $h_2$ for each of the $S=25$ design points for $a=0$ (a) and $a=0.5$ (b). Panel (c) shows the addition of new points in the control variable input space and panel (d) shows the evolution of the EQI metric both corresponding to the algorithm run from panel (b).}
\label{seq:design:non_aggressive}
\end{figure}

In this section we present the results from the illustrative example (Section~\ref{sec:Optimsation:illustrative_example}) of running the ``aggressive" version of the algorithm (searching for points which dominate at least one current solution) for the first 10 steps and then running the ``non-aggressive" version (expand the set of non-dominated solutions) for a second set of 10 steps, see Figure~\ref{seq:design:non_aggressive}. It is clear that a greater proportion of the points lie on the Pareto front compared to running just the aggressive algorithm. Depending on computational budget and the goal of the study, one may wish to run the aggressive algorithm for a few steps to get close to the true Pareto front quickly, and then switch to the non-aggressive algorithm to add more points to the Pareto front.

\figref{MOEEQI_evolution} shows the evolution of the MO-E-EQI metric over 20 iterations. There is a jump at iteration 11 as the integration area has switched to include the shaded area in \figref{fig:Pareto-set} so that the algorithm can fill in the gaps in the observations.

\section{Ground truth}\label{app:truth}

\begin{figure}[!ht]
\centering
\begin{subfigure}{.5\textwidth}
\begin{center}
\includegraphics[width=\textwidth]{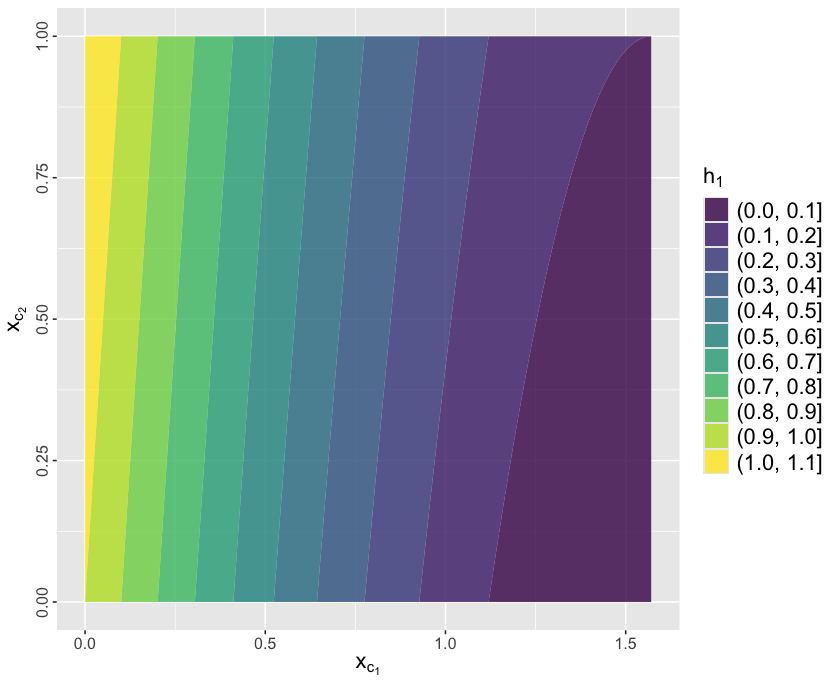}
\end{center}
\caption{}
\end{subfigure}%
\begin{subfigure}{.5\textwidth}
\begin{center}
\includegraphics[width=\textwidth]{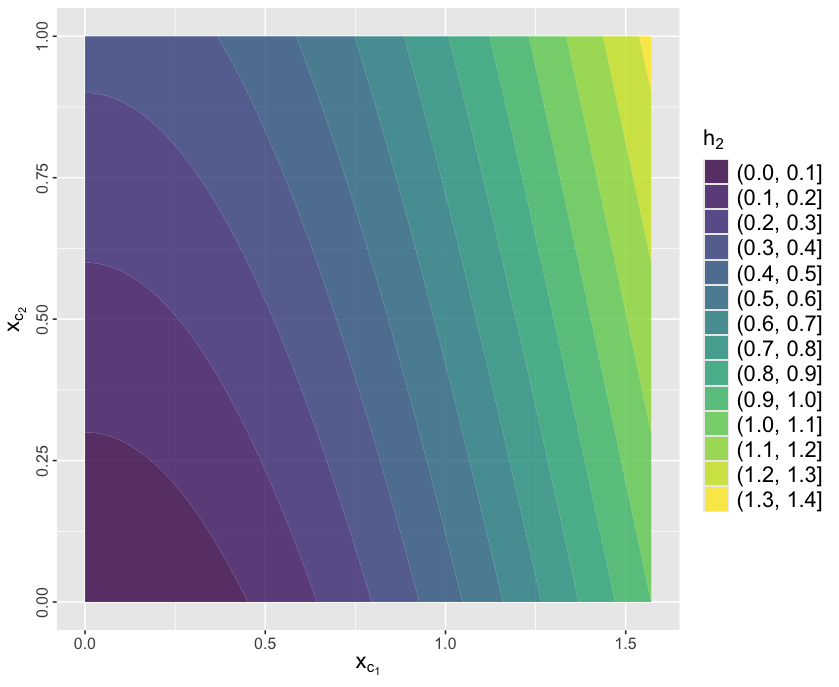}
\end{center}
\caption{}
\end{subfigure}%
\caption{
Response contours of $h_1$ and $h_2$ in \eqref{eq:toy} with environmental variables' noise integrated out.}
\label{fig:ground_truth}
\end{figure}

In the example from Section~\ref{sec:Optimsation:illustrative_example}, both functions $h_1$ and $h_2$ have noise resulting from environmental variables. \figref{fig:ground_truth} presents the response contours without environmental variables:  
$h_1(\bx_c) = 1-\sin(x_{c_1})+\frac{x_{c_2}}{10}$ and 
$h_2(\bx_c) = 1-\cos(x_{c_1})+\frac{x_{c_2}}{3}.$
In this case, these functions would also result from integrating out the noise as the distributions for both environmental variables are symmetric and centered at 0.

\end{document}